\documentclass[acmsmall,screen]{acmart}

\usepackage{algorithmic}
\usepackage{graphicx}
\usepackage{textcomp}
\usepackage{subfig}
\usepackage[utf8x]{inputenc}
\usepackage{multirow}
\usepackage{balance}
\usepackage{booktabs}
\usepackage{xspace}
\usepackage[colorlinks,linkcolor=blue,citecolor=blue]{hyperref}
\usepackage{tabularx}
\newcolumntype{R}{>{\raggedleft\arraybackslash}X}
\usepackage[normalem]{ulem}

\usepackage{listings}
\definecolor{light-gray}{gray}{0.95}
\definecolor{pgreen}{RGB}{5,205,107}
\definecolor{pblue}{RGB}{2,154,223}
\makeatletter
\newenvironment{btHighlight}[1][]
{\begingroup\tikzset{bt@Highlight@par/.style={#1}}\begin{lrbox}{\@tempboxa}}
{\end{lrbox}\bt@HL@box[bt@Highlight@par]{\@tempboxa}\endgroup}

\newcommand\btHL[1][]{%
  \begin{btHighlight}[#1]\bgroup\aftergroup\bt@HL@endenv%
}
\def\bt@HL@endenv{%
  \end{btHighlight}%
  \egroup
}
\newcommand{\bt@HL@box}[2][]{%
  \tikz[#1]{%
    \pgfpathrectangle{\pgfpoint{1pt}{0pt}}{\pgfpoint{\wd #2}{\ht #2}}%
    \pgfusepath{use as bounding box}%
    \node[anchor=base west, fill=orange!30,outer sep=0pt,inner xsep=1pt, inner ysep=0pt, rounded corners=3pt, minimum height=\ht\strutbox+1pt,#1]{\raisebox{1pt}{\strut}\strut\usebox{#2}};
  }%
}
\makeatother

\lstdefinestyle{log}{
    keywordstyle=\ttfamily,
    moredelim=**[is][{\btHL[fill=green!30]}]{**}{**}
}

\AtBeginDocument{%
}

\usepackage[framemethod=tikz]{mdframed}
\mdfdefinestyle{mpdframe}{
    frametitlebackgroundcolor   =black!15,
    frametitlerule              =true,
    roundcorner                 =1pt,
    middlelinewidth             =1pt,
    skipabove                   =\baselineskip,
    innermargin                 =0.1cm,
    innerleftmargin             =0.1cm,
    innerrightmargin            =0.1cm,
    innertopmargin              =0.1cm,
    innerbottommargin           =0.1cm
}

\newcommand\chaoseth{{\sc ChaosETH}\xspace}
\newcommand\chaosethbf{{\sc \textbf{ChaosETH}}\xspace}

\AtBeginDocument{%
  \providecommand\BibTeX{{%
    \normalfont B\kern-0.5em{\scshape i\kern-0.25em b}\kern-0.8em\TeX}}}

\setcopyright{acmcopyright}
\copyrightyear{2023}
\acmYear{2023}
\acmDOI{XXXXXXX.XXXXXXX}

\acmJournal{DLT}

\DeclareUnicodeCharacter{221A}{$\sqrt{ }$}

\begin{document}

\title{Chaos Engineering of Ethereum Blockchain Clients}

\author{Long Zhang}
\email{longz@kth.se}
\orcid{0000-0002-7211-3894}
\affiliation{%
  \institution{Electrolux}
  \streetaddress{S:t Göransgatan 143}
  \city{Stockholm}
  \country{Sweden}
  \postcode{112 17}
}
\affiliation{%
  \institution{KTH Royal Institute of Technology}
  \streetaddress{Lindstedtsvägen 3}
  \city{Stockholm}
  \country{Sweden}
  \postcode{100 44}
}

\author{Javier Ron}
\email{javierro@kth.se}
\orcid{0000-0001-6988-3102}
\author{Benoit Baudry}
\orcid{0000-0002-4015-4640}
\email{baudry@kth.se}
\author{Martin Monperrus}
\email{monperrus@kth.se}
\orcid{0000-0003-3505-3383}
\affiliation{%
  \institution{KTH Royal Institute of Technology}
  \streetaddress{Lindstedtsvägen 3}
  \city{Stockholm}
  \country{Sweden}
  \postcode{100 44}
}

\renewcommand{\shortauthors}{Zhang et al.}

\begin{abstract}
In this paper, we present \chaoseth, a chaos engineering approach for resilience assessment of Ethereum blockchain clients. \chaoseth operates in the following manner: First, it monitors Ethereum clients to determine their normal behavior. Then, it injects system call invocation errors into one single Ethereum client at a time, and observes the behavior resulting from perturbation. Finally, \chaoseth compares the behavior recorded before, during, and after perturbation to assess the impact of the injected system call invocation errors. The experiments are performed on the two most popular Ethereum client implementations: GoEthereum and Nethermind. We assess the impact of 22 different system call errors on those Ethereum clients with respect to 15 application-level metrics. Our results reveal a broad spectrum of resilience characteristics of Ethereum clients w.r.t. system call invocation errors, ranging from direct crashes to full resilience. The experiments clearly demonstrate the feasibility of applying chaos engineering principles to blockchain systems.
\end{abstract}



\maketitle

\section{Introduction}
\label{sec:introduction}

Starting from the Bitcoin system \cite{BitcoinPaper}, blockchain techniques have drawn much attention because of their strong trustworthiness and reliability \cite{Tang:SACMAT19:IoTPassport, Cai:TDSC2021:TowardsCrowdsensingSystems}. Ethereum is one of the most popular blockchain platforms \cite{Chen:SurveyOnEthSecurity} that supports both cryptocurrencies and decentralized finance \cite{werner2021sok}. In order to join the Ethereum distributed network, participants run an Ethereum client in their local environment. Then, the network is composed of all clients interacting together. There are multiple implementations of Ethereum clients \cite{aumasson2021SecurityReview}, written in different languages, all implementing the same specification and protocol \cite{ETHYellowPaper}.

The consensus protocol is designed to provide resilience against high-level malfunctions. Yet, low-level bugs in different clients may affect the network as a whole. Thus, in addition to protocol-level resilience, it is essential to understand and improve the code-level resilience of the Ethereum client implementations.
Ethereum clients indeed have many reasons to malfunction sometimes, such as operating system overload, memory errors, network partition, etc.
Meanwhile, recent surveys \cite{Natella:2016:FISurvey, Xiang:Availability_resilience_and_fault_tolerance_of_internet_and_distributed_computing_systems} stress the lack of work on assessing the resilience of blockchain clients and Ethereum.

Chaos engineering is a novel methodology to assess and improve the error-handling mechanisms of software systems \cite{Basiri:Chaos_Engineering:IEEESoftware}. To perform chaos engineering, developers actively inject failures in a system in production, in a controlled manner. This allows them to compare the behavior observed during fault injection with the system's normal behavior \cite{Chaos_Engineering_Book:OReilly}. The system can be considered resilient If these behaviors are similar. However, discrepancies between them can indicate the presence of resilience issues. In the context of an Ethereum client, it is very challenging to predict and test offline the problems that an Ethereum client will meet after it is deployed in production because it is impossible to reproduce the actual Ethereum blockchain evolution at scale. 
Chaos engineering is fit to address this challenge, as it triggers an Ethereum client's error-handling code while the client is executing in a full-fledged production environment, while downloading, sharing, and verifying the main blockchain.

In this paper, we present the design and implementation of a novel chaos engineering methodology called \chaoseth. \chaoseth analyzes an Ethereum client resilience by perturbing its system call invocations  in production. 
We focus on system calls as they are known to be appropriate to capture the behavior of a software system \cite{forrest:sense-of-self,ghavamnia2022c2c}.
The implementation of \chaoseth is applicable to any Ethereum client, regardless of the programming language used. The resilience assessment is based on advanced monitoring that captures the steady state of an Ethereum client. As a result, \chaoseth is able to identify resilience strengths and weaknesses in Ethereum clients.

\chaoseth is evaluated by conducting chaos engineering experiments on GoEthereum v1.10.25 and Nethermind 1.14.5. During the experiments, \chaoseth perturbs GoEthereum and Nethermind while they interact with other Ethereum nodes all over the world.
The results show that 1) \chaoseth successfully conducts chaos engineering experiments using different error models, and 2) \chaoseth successfully identifies different degrees of resilience with respect to system call invocation errors. We also present an original resilience benchmarking using 4 common error models (defined in \autoref{sec:experiment-protocol-benchmarking}) for a sound comparison of their resilience.
The core novelty of our work is to perform reliability assessment of blockchain systems under production conditions. To our knowledge, we are the first to deeply study the usage of chaos engineering for blockchain systems \cite{Huang:SurveyOnBlockchains}.

In summary, the main contributions of this paper are:

\begin{itemize}

\item A novel methodology to perform chaos engineering on blockchain clients for resilience assessment. This assessment is done directly in production to avoid the core limitations of blockchain test networks. The methodology is fully implemented in a tool called \chaoseth dedicated to Ethereum.

\item An empirical evaluation of the resilience of GoEthereum v1.10.25 and Nethermind 1.14.5 in production, with respect to realistic error models at the system call level. The chaos engineering experiments highlight strengths (including resilience) and weaknesses (including crashes) for both clients, which is valuable knowledge for the Ethereum community.

\item An empirical and sound resilience benchmarking of GoEthereum versus Nethermind with respect to $4$ common error models. We compare the differences in error handling capabilities between these clients against the same system call invocation errors, and show that no client is consistently more resilient.

\item An implementation of \chaoseth that supports full error injection for system call invocations in an Ethereum client. The system is publicly available for future research (\url{https://github.com/KTH/royal-chaos/tree/master/chaoseth}).

\end{itemize}

The rest of the paper is structured as follows: \autoref{sec:background} introduces the background knowledge of blockchain techniques and chaos engineering. \autoref{sec:design} and \autoref{sec:experimentation} present the design and evaluation of \chaoseth. \autoref{sec:discussion} discusses threats to validity, ethical considerations, and the applicability of \chaoseth to other Ethereum clients, and \autoref{sec:conclusion} concludes the paper.

\section{Background}\label{sec:background}

\subsection{Blockchain Nodes}
Blockchain is the core technology behind the success of cryptocurrencies such as Bitcoin \cite{BitcoinPaper}.
A blockchain is a decentralized distributed ledger for recording consensual information \cite{Wu:IEEENetwork2020:BlockEdge}. The information is saved in a sequence of ``blocks'' which are shared by all of the participants on a blockchain network.

In this work, we focus on Ethereum \cite{ETHYellowPaper}, an open-source blockchain system that supports the execution of smart contracts. A smart contract is a program that is persisted on-chain, and is executable by every user on the network using the so-called Ethereum Virtual Machine (EVM). 
As introduced in Ethereum's official documentation \cite{EthereumDocs}, an Ethereum client is an implementation of Ethereum that 1) verifies blocks by executing contracts, and 2) shares them with other peers over the Internet. 
There exist different Ethereum clients that implement an EVM in different programming languages, such as \texttt{go-ethereum} and \texttt{nethermind}.
The host that runs a given Ethereum client is defined as a ``node'', there are hundreds of thousands of such nodes at the time of writing.

\begin{figure*}
\centering
\includegraphics[width=\textwidth]{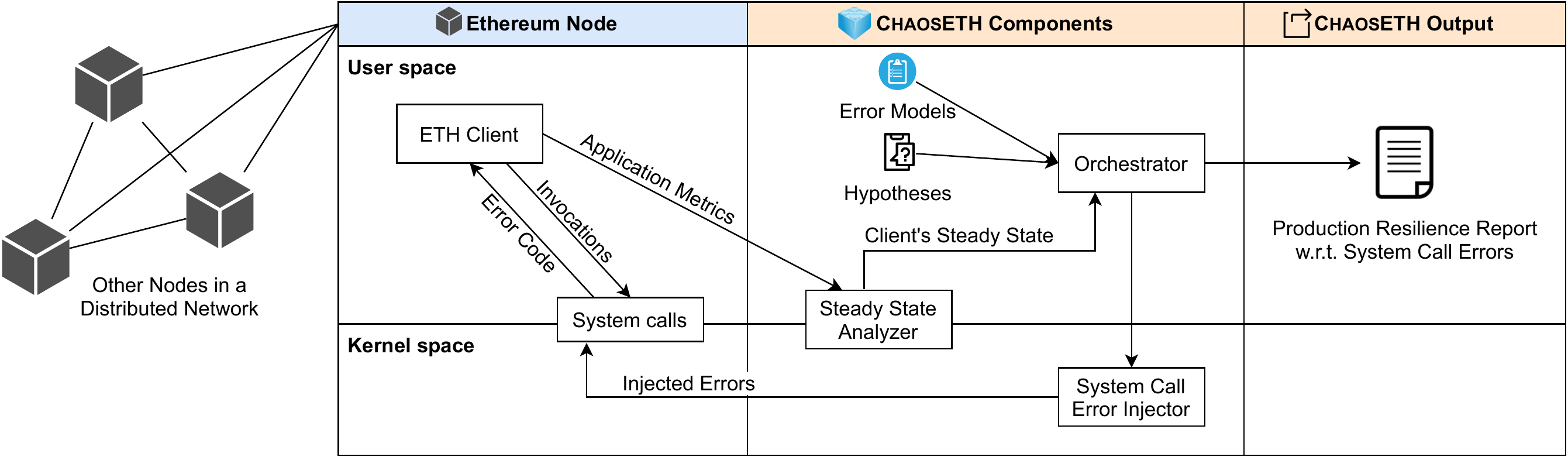}
\caption{The components of \chaoseth}\label{fig:components}
\end{figure*}

\subsection{Chaos Engineering}\label{sec:background-ce}

Chaos engineering is a software reliability methodology to assess error-handling capabilities by actively triggering failures in production \cite{Principles_Of_Chaos_Engineering}. For example, Netflix uses ChaosMonkey, which randomly shuts down servers in production to verify that the whole system is able to recover from such a failure scenario \cite{Basiri:Chaos_Engineering:IEEESoftware}. Chaos engineering is a unique complement to other software reliability techniques because it is done after a system is deployed in production, where the running environment is not as controlled as during testing or staging \cite{Zhang:ChaosMachine}.

The application of chaos engineering is based on a series of steps \cite{Chaos_Engineering_Book:OReilly}. First, the target system's steady state needs to be defined. The steady state is a set of observable metrics which characterize the system's normal behavior. For instance, the number of video streams opened by users per second could be used as a steady state metric for Netflix \cite{Basiri:Chaos_Engineering:IEEESoftware}. After defining the steady state, developers set hypotheses that describe the expected behavior of the target system during a chaos engineering experiment. In this context, a chaos engineering experiment is an execution period during which failures are injected into the target system. If a hypothesis holds, it means that the target system behaves as expected when a certain failure happens in production.
Otherwise, if the hypothesis is falsified, actions need to be taken based on the knowledge learned from the chaos engineering experiment. Finally, as the experiment is performed in production environments, developers have to limit the side effects on user experience caused by the experiment, this is also called ``blast radius control''. For example, developers could utilize containerization techniques to isolate the chaos engineering experiment targets \cite{Simonsson:ChaosOrca:FGCS2021}.

In this paper, we deploy the full chaos engineering methodology in the context of blockchain systems. We define the steady state, set up a fault injection model, formalize hypotheses and conduct chaos engineering for Ethereum blockchain nodes.

\subsection{Related Work}\label{sec:related-work}

\subsubsection{Blockchain Dependability}

There are several surveys that focus on blockchain \cite{Huang:SurveyOnBlockchains, Fan:Access2020:PerformanceEvaluationSurvey, Chen:SurveyOnEthSecurity, Wang:EthSmartContractSecurityResearch, Atzei:SurveyOfAttacksOnSmartContracts}. For example, Huang et al. \cite{Huang:SurveyOnBlockchains} conducted a survey of the state of the art on blockchains including the theories, models, and tools. Based on these surveys, we notice that there is limited research on using fault injection techniques to evaluate the reliability of blockchain systems. Most of the recent works in this direction is about fuzzing smart contracts. To our knowledge, there is no work on perturbing the runtime environment of  blockchain node, as is done by \chaoseth.

Regarding performance, the existing research works focus on different levels, including EVM opcode level \cite{Aldweesh:AICCSA2018:PBenchmarkingForOpcodes, Aldweesh:2019:OpBench}, smart contract level \cite{Aldweesh:edcc2018:PerformanceBenchmarking}, consensus algorithms level \cite{Hao:IV2018, Ahmad:ICOIN2021}, blockchain system level \cite{Dinh:SIGMOD2017:BLOCKBENCH, Dabbagh:IICAIET2020:PerformanceAnalysis, Ampel:ISI2019, cortesgoicoechea2020ResourceAnalysis, Armknecht:SACMAT2021}, and blockchain-based application level \cite{Truong:BenchmarkingBlockchainInteractionsInMECSS:2019}.
Compared with these works, \chaoseth does not focus on performance evaluation of blockchain systems. Instead, performance-related metrics such as memory usage are used by \chaoseth to evaluate the side effects caused by system call invocation errors.
The most related work is done by Dinh et al. \cite{Dinh:SIGMOD2017:BLOCKBENCH} who invented BlockBench, a framework that evaluates a private blockchain's performance using its throughput, latency, scalability, and fault-tolerance capability as indicators. BlockBench investigates how Byzantine failures affect a blockchain system's throughput and latency by crashing some nodes, injecting network delays, and corrupting messages among different nodes.
Compared with BlockBench, \chaoseth assesses a blockchain client using a public blockchain network, which means developers do not need to have the full control of the whole decentralized system. The failure models are also different between these two works.

Regarding reliability analysis and improvement,
Seol et al. \cite{Seol:2018} implemented a blockchain analytics engine for assessing the dependability of off-chain file systems.
Sousa et al. \cite{Sousa:DSN2018:BFTSmartOrderingService} designed a Byzantine fault-tolerant ordering service for Hyperledger Fabric.
Zhang et al. \cite{Zhang:ICBC2018} designed LedgerGuard, a mechanism that keeps the integrity of a ledger via corrupted blocks detection and recovery.
Liu et al. \cite{Liu:HotICN2018} proposed an evaluation methodology that applies a continuous-time Markov chain model for blockchain-based IoT applications.
It has been proposed to use modeling techniques to study blockchain reliability. For instance, Melo et al. \cite{Melo:ISCC2018} proposed a modeling methodology that evaluates reliability and availability of a blockchain-as-a-service environment. Kancharla et al. \cite{Kancharla:2020:HybridChainDependability, Kancharla:2020:SlimChainDependability} applied simulation methods to demonstrate the dependability of the proposed hybrid blockchain and slim blockchain.
González et al. \cite{gonzalez:hal-01653986} categorized different fault injection techniques and discussed the possibilities to apply them in blockchain-based applications resilience assessment.
Instead of using simulations or focusing on blockchain-based applications, \chaoseth uses real Internet traffic to assess the resilience of Ethereum clients in production.

Regarding security,
research has been done on analyzing vulnerabilities that are located in smart contracts \cite{Correas:StaticProfilingAndOptimizationOfSmartContracts, Alkhalifah:DetectAndPreventReentrancyAttacks, Ashraf:GasFuzzer, Fu:EVMFuzzer:FSE2019, Hajdu:2020:FIforFaultySmartContracts, Wang:2020:ContraMaster, Shlomi:2021:EtherProv}. For example, Fu et al. \cite{Fu:EVMFuzzer:FSE2019} designed EVMFuzzer, a framework that generates contracts via a set of predefined mutators to find security bugs in different EVM implementations.
Zhang et al. \cite{Zhang:USENIX2020:TXSPECTOR} presented TxSpector, a logic-driven approach for detecting attacks in Ethereum transactions at the bytecode level.
Aumasson et al. \cite{aumasson2021SecurityReview} reviewed four Ethereum clients from a security perspective. Their work consists of a benchmarking methodology as all of the four clients are evaluated using the same set of security problems. However it is different from this paper because \chaoseth focuses on resilience benchmarking instead of security.

Moreover, none of these solutions support the specification and execution of fault injection experiments directly in a production-like environment. \chaoseth is the first methodology that provides developers with a systematic way to learn how their Ethereum client implementations react to different system call invocation errors in production. 

\subsubsection{Chaos Engineering}
Basiri et al. \cite{Basiri:Chaos_Engineering:IEEESoftware} presented the principles of chaos engineering in 2016. 
The earliest known chaos engineering tool is called the `Chaos Monkey' \cite{Basiri:Chaos_Engineering:IEEESoftware}, which has never been deployed to Ethereum, as opposed to the `Bored Ape'\cite{BoredApe}.
Zhang et al. \cite{Zhang:ChaosMachine} designed ChaosMachine, a tool that conducts chaos engineering experiments at the try-catch level for Java applications.
Jernberg et al. \cite{Jernberg:GettingStartedWithCE:ESEM2020} designed a chaos engineering framework based on the literature and a tool survey, and validated the framework in a real commercial web system.
Simonsson et al. \cite{Simonsson:ChaosOrca:FGCS2021} proposed ChaosOrca, a chaos engineering system that injects system call errors for dockerized applications.
Hernández-Serrato et al. \cite{HernandezSerrato:ApplyMLWithCE:ISSREW2020} discussed the possibilities to apply machine learning techniques for improving chaos engineering experiments.
Ikeuchi et al. \cite{Ikeuchi:AutomaticFailureRecoveryInICTSystems:ICDCS2020} proposed a framework for learning a recovery policy using deep reinforcement learning and chaos engineering techniques.
Chaos engineering is also applicable in the field of security. Torkura et al. \cite{Torkura:CloudStrike:2020} proposed CloudStrike, a tool that focuses on injecting failures which impact security i.e. integrity, confidentiality and availability.
Regarding human factors in chaos enginering, Canonico et al. \cite{Canonico:Human-AIPartnershipsForCE:ICSEW2020} discussed what aspects of AI would be used to make a system more resilient to perturbations and the results of these findings against existing chaos engineering approaches.

The only related work that combines chaos engineering and blockchains is by Sondhi et al. \cite{Sondhi:arXiv2021CEforConsensusAlgorithms}. They apply chaos engineering to evaluate the performance of different consensus algorithms in permissioned blockchains. Their perturbation models were designed for representing network failures and message corruptions.
In comparison, \chaoseth considers an entirely different failure model: errors at the level of system call invocations. This failure model captures the problems that may happen on the node's operating system and not on the network. To our knowledge, our methodology of end-to-end chaos engineering for blockchain systems is novel.

\section{Design of \chaosethbf}\label{sec:design}

\chaoseth assesses an Ethereum client's resilience within a real production environment. This section introduces the motivation behind \chaoseth, as well as its design and implementation.

\subsection{Motivation}\label{sec:motivation}

An Ethereum client is always executed on top of an operating system. The operating system is responsible for providing the Ethereum client access with critical resources such as network and storage, and it does so by means of system calls. 
For illustration, during a short 1-minute observation period of the GoEthereum client, we observed more than one million system calls ($1,128,215$) that cover $36$ different types such as \texttt{read} and \texttt{write}. Although the client is behaving normally, the system call invocations are not all successful: $14,640$  system call invocations fail during the observation with $9$ different kinds of error codes within the same minute-long observation period.
The fact that GoEthereum can stay up and running even if some of the system call invocations fail shows that GoEthereum is equipped with certain error-handling mechanisms.

As a rule of thumb, both developers and users of Ethereum clients seek robustness.
\chaoseth is meant to help them to better understand and potentially improve this robustness.

\subsection{Challenges of Applying Chaos Engineering to Blockchain Systems}\label{sec:challenges}

In order to apply chaos engineering to blockchain systems, the following challenges have to be addressed.
First of all, it is a challenging task to select appropriate metrics for defining the steady state prior to chaos engineering experiments. Even though most clients expose many metrics at different levels, it is unclear which ones are useful for reasoning about the behavior under fault injection.
Secondly, it is complex to design appropriate fault injection models. Even if one only considers system call invocation errors, there still exist hundreds of system call types and error codes. Given a limited experiment time, focused and realistic fault injection models have to be guaranteed.
Last but not the least, it is challenging to assess the correctness of a blockchain system since the blockchain is always changing in production, one block at a time.
The approach presented in this paper tackles those three challenges.

\subsection{Overview}

In order to achieve the goal of assessing the resilience of blockchain clients, our core requirements are:
1) finding resilience problems related to production environments, hence \chaoseth should work on deployed clients,
2) finding resilience problems related to real data, hence \chaoseth should operate with the real Ethereum blockchain,
and 3) being applicable to different implementations of Ethereum so that benchmarking \cite{Kanoun:Dependability_Benchmarking_for_Computer_Systems} can be conducted, for the community to identify the most robust clients.
Our contribution, \chaoseth, satisfies those three requirements.

\chaoseth is composed of three components, and produces resilience reports for developers, as summarized in \autoref{fig:components}. 
Recall that Ethereum is a distributed network with many nodes running together. \chaoseth operates on one single node (signified by the column titled Ethereum Node in \autoref{fig:components}). In this specific Ethereum node, \chaoseth is attached to the client process during its execution (say the GoEthereum process) .
The steady state analyzer collects monitoring metrics and infers the system's steady state. We will present this in more detail in \autoref{sec:steady-state-analyzer}.
The system call error injector injects different error codes into system call invocations in a controlled manner, see \autoref{sec:injector}.
With the help of the orchestrator, introduced in \autoref{sec:orchestrator}, \chaoseth systematically conducts chaos engineering experiments.
Finally, \chaoseth produces a resilience report with respect to system call errors for developers. We discuss the resilience report in \autoref{sec:reports}.

\subsection{Components of \chaosethbf}

\subsubsection{Steady State Analyzer}\label{sec:steady-state-analyzer}

As introduced in \autoref{sec:background-ce}, defining the steady state of the Ethereum client under study is essential for a chaos engineering experiment. 
Per the state of the art of observability, all Ethereum clients provide some monitoring capabilities.
In \chaoseth, the steady state analyzer collects behavior-related metrics directly from the monitoring component provided by an Ethereum client\footnote{See \url{https://geth.ethereum.org/docs/monitoring/metrics} and \url{https://docs.nethermind.io/nethermind/ethereum-client/metrics}.}. For instance, the latest version of the GoEthereum client, v1.10.25, exposes more than $400$ different metrics that describe the runtime status of the client. By reusing these pre-existing mechanisms, there is a guarantee that no extra monitoring overhead is introduced.

Recall the first challenge mentioned in \autoref{sec:challenges}, not all the provided metrics are appropriate for describing the steady state. For instance, the highest block number is not an ideal metric for steady state inference because this metric is determined by the whole blockchain network instead of the client itself. In order to address this challenge, the steady state analyzer first needs developers to define a set of metrics from the client's monitoring module. Then, the analyzer conducts statistical analysis on the selected metrics and confirms if a metric is statistically stable for describing the client's steady state.

For the selected metrics, the monitoring is done per `monitoring interval', which is a period of time during which data is collected and persisted. The monitoring interval is set by developers and is an engineering trade-off for developers between overhead and precision \cite{Zhang:phoebe}.
A shorter monitoring interval may produce a better picture of the steady state, but also means a larger amount of monitoring data and a higher monitoring overhead. 
A typical monitoring interval is 15 seconds.

For the steady state inference task, we first configure a `monitoring epoch' in the analyzer. A monitoring epoch is defined as a sequence of contiguous monitoring intervals. For example, if the monitoring epoch is $3600$ seconds and the monitoring interval is $15$ seconds, the analyzer records $3600 / 15 = 240$ data points for each metric. For a given configuration, the analyzer samples all metrics over two monitoring epochs. Then, for each metric, the samples from the two monitoring epochs are compared to each other using the Mann-Whitney U test. This statistical test is used to determine if the probability distribution of the two epochs is different. The stable metrics are those whose distributions over the two epochs are not statistically different, and are selected to describe the client's steady state.

After inferring the client's steady state by means of metric distributions, \chaoseth can conduct chaos engineering experiments, where the core idea is to compare the metric distributions under error injection against the reference distributions.

\subsubsection{System Call Error Injector}
\label{sec:injector}
In order to conduct chaos engineering experiments, we need to inject errors.
In this paper, we focus on system call errors and design a system call error injector accordingly.
When the injector is activated, it listens to system call invocation events, and it replaces the original return code with an error code. In \chaoseth, the error injector uses an error injection model defined as a triple \texttt{(s, e, r)}: it means that system call \texttt{s} is injected with error code \texttt{e} under the error rate \texttt{r}$\in [0,1]$. The error rate is the probability of replacing the return code by error code \texttt{e} when system call \texttt{s} is invoked at runime.
For example, an error injection model \texttt{(read, EAGAIN, 0.5)} means that every time a \texttt{read} system call is invoked, there is a $50\%$ probability that a successful return code is replaced with error code \texttt{EAGAIN}, which represents that the resource being read is temporarily unavailable \cite{errorno-manpage}.

Depending on the type of system call to perturb and the error rates, the error scenario is more or less severe. The higher the error rate in an error injection model, the more frequently such errors are injected into the target client, which means that the client is subject to a higher resilience pressure.

Considering the second challenge mentioned in \autoref{sec:challenges}, \chaoseth aims at focusing on realistic errors. For this, we follow the state of the work by Zhang et al. \cite{Zhang:phoebe}. According to their methodology realistic fault models should fulfill two characteristics: they represent real production failure scenarios, and happen frequently enough so that developers are able to analyze the behavior during a chaos engineering experiment. Consequently, in \chaoseth our fault models are realistic because:
1) they use the same error code as system call invocation errors that are observed in the field, 
2) the error rate is amplified so that more invocation errors can be observed in a short period of time and trigger resilience issues.

\subsubsection{Experiment Orchestrator}
\label{sec:orchestrator}

The experiment orchestrator communicates with the steady state analyzer and the system call error injector to conduct chaos engineering experiments. The orchestrator attaches these two components to the operating system kernel before the Ethereum client starts. Then it activates the error injector according to the experiment configuration. An experiment configuration defines the duration of an experiment, and the corresponding error injection model to be used.

The orchestrator divides each chaos engineering experiment into five phases: 1) the warm-up phase, 2) the pre-checking phase, 3) the error injection phase, 4) the recovery phase, and 5) the validation phase. During the warm-up phase, the target client runs until it reaches its steady state. During the pre-checking phase, the steady state analyzer is activated to monitor the client and to check if the client has already reached its steady state. During the error injection phase, both the steady state analyzer and the error injector are turned on so that system call invocation errors are injected according to the given error injection model. After the error injection phase is done, the orchestrator gives the target client time to recover during a recovery phase. After this recovery phase, the validation phase takes place during which the orchestrator monitors the client's state again. The observed behavior during this validation phase is compared with the previously inferred steady state using the Mann–Whitney U test. If the target client fails to recover back to its steady state after the validation phase, it means that the injected errors have caused an adverse side effect on the client. This can signify to the developers that the client's error handling mechanisms may have failed to recover from the injected error.

\subsubsection{Chaos Engineering Hypotheses}
\label{sec:reports}

After a chaos engineering experiment, the orchestrator analyzes the client's behavior during the experiment and generates a resilience report for developers. The report presents to what extent the implementation of the Ethereum client under evaluation is resilient to the injected system call invocation errors. As introduced in \autoref{sec:background-ce}, the client's resilience is assessed by validating chaos engineering hypotheses. \chaoseth uses three hypotheses about how a client reacts when errors are injected using one specific error model.

\paragraph{Non-Crash hypothesis ($H_N$)}
The non-crash resilience hypothesis holds if the injected system call invocation errors do not crash the Ethereum client and the process remains alive.

\paragraph{Observability hypothesis ($H_O$)}
The observability hypothesis for a metric $m$ holds if $m$ is influenced by error injections according to an error model.

\paragraph{Recovery hypothesis ($H_R$)}
The recovery hypothesis is valid if the client is able to recover to its steady state after stopping the injection of system call invocation errors.

For example, let us assume that the system call error injector follows the error model \texttt{(read, EAGAIN, 0.5)} to conduct the experiment. The target client does not crash during fault injection. However, the success rate of reading  data from disk drops significantly during fault injection because the client fails at invoking the \texttt{read} system call. Furthermore, the success rate of reading data gets back to its steady state again after the experiment stops. In this case, \chaoseth reports that all of the three hypotheses $H_N$, $H_O$ and $H_R$ are validated, meaning that the target client is non-crashing, observable, and resilient with respect to error model \texttt{(read, EAGAIN, 0.5)}.
By validating or falsifying the hypotheses above, developers learn more about the Ethereum client's resilience with respect to different types of system call invocation errors. Such information is helpful for developers to prioritize their work on improving error handling.

\subsubsection{Implementation}

The system call monitor and system call error injector are based on the Phoebe framework \cite{Zhang:phoebe} which implements them with the help of the eBPF module (extended Berkeley Packet Filter). More specifically, the monitor and error injector register their BPF programs to the \texttt{sys\_enter} and \texttt{sys\_exit} events. When there is an error that needs to be injected, the error injector calls the BPF helper function \texttt{bpf\_override\_return} to replace the original return code with the error code. The experiment orchestrator is implemented in Python. The source code of \chaoseth is publicly-available at \url{https://github.com/KTH/royal-chaos/tree/master/chaoseth}.

\section{Experimentation}\label{sec:experimentation}

\subsection{Overview}
\newcommand\rqsteadystate{What is the feasibility of steady state characterization of Ethereum clients with \chaoseth?}

\newcommand\rqresilience{What are \chaoseth's resilience properties actually verified in the Ethereum clients under study?}

\newcommand\rqbenchmarking{What are the resilience differences among the considered Ethereum clients with respect to system call errors?}

In order to evaluate the effectiveness of \chaoseth, we propose the following three research questions.

\begin{itemize}
    \item RQ1: \rqsteadystate
    \item RQ2: \rqresilience
    \item RQ3: \rqbenchmarking
\end{itemize}

\begin{figure*}
\centering
\includegraphics[width=\textwidth]{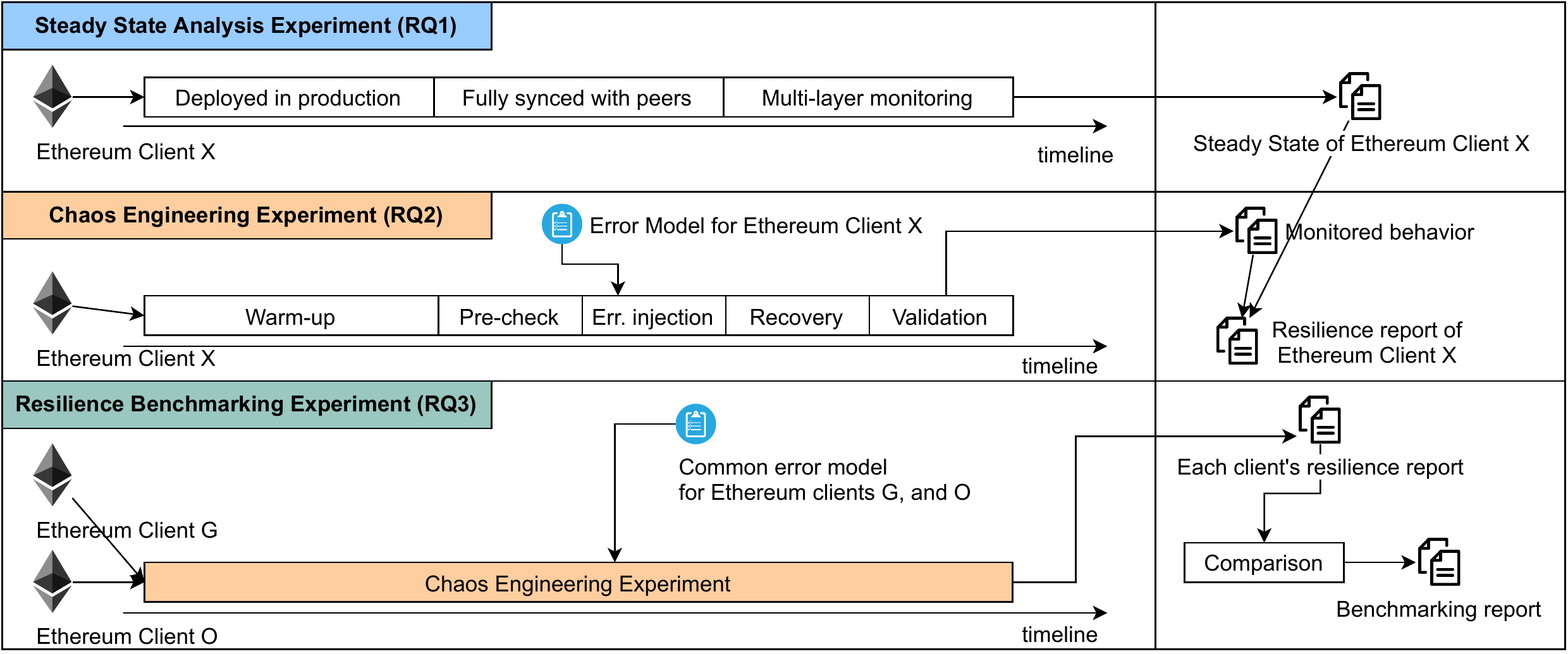}
\caption{Experimental Workflow}\label{fig:experiments}
\end{figure*}

All of the research questions are answered quantitatively by conducting experiments on the selected Ethereum client implementations. To maximize the relevance and impact of our research, we select the most popular Ethereum clients based on the statistical information about Ethereum's main network published at \url{ethernodes.org}. At the time of writing, the most popular two Ethereum clients are GoEthereum and Nethermind. Thus, we select GoEthereum v1.10.25 and Nethermind 1.14.5 for the experiments. Our experimental workflow is summarized in \autoref{fig:experiments}, and we discuss it in details in \autoref{sec:experiment-protocol}. The rest of the section is dedicated to the presentation of the results.

\subsection{Experiment Protocol}\label{sec:experiment-protocol}
To assess the resilience of the two considered Ethereum clients, we conduct three different categories of experiments: steady state analysis experiments, chaos engineering experiments, both performed individually on each client; and resilience benchmarking experiments, performed collectively on both of the clients. Both Ethereum clients run with a production configuration in bare-metal servers. The GoEthereum client runs in a server with specs: Intel(R) Core(TM) i9-10980XE / 128 GB RAM / 2TB NVMe SSD with Ubuntu 18.04.6. The Nethermind client runs in a server with Intel(R) Core(TM) i9-10900K / 48 GB RAM / 2TB SATA SSD with Ubuntu 20.04.5.

\subsubsection{Steady State Analysis Experiment}
\label{sec:experiment-protocol-steadystate}

To answer RQ1, we conduct a steady state analysis experiment on each client.
As introduced in \autoref{sec:background-ce}, characterizing an Ethereum client's steady state is a prerequisite for a chaos engineering experiment. Conducting steady state analysis gives developers a clear understanding of how the Ethereum client behaves in a normal production condition.

The first row of \autoref{fig:experiments} illustrates our methodology to model the steady state. Each Ethereum client (GoEthereum or Nethermind), is started up using the options recommended in their respective documentation. The clients are started in active synchronization mode, which consists in downloading and verifying the blocks from other peers on the Internet. After starting up the client, we wait for the client to finish synchronizing the existing blocks.

After a client reaches a synced state, the main job of the client is to verify newly generated blocks and to share all the blocks with other peers. From this point, the steady state analyzer is attached to the client, with a monitoring interval of 15 seconds and a monitoring epoch of 5 hours, per our conceptual framework detailed in \autoref{sec:steady-state-analyzer}.
These values ensure that a sufficient number of data points ($5\times60\times60\div15=1200$ points) are available for capturing the distribution of each metric.

A steady state analysis experiment takes all of a client's exposed metrics as input and outputs each metric's probability distribution. The metrics that meet the following two criteria are selected for chaos engineering experiments: 1) being an active non-zero metric  (some metrics are always zero if a feature is not turned on), and 2) being statistically stable to describe the client's steady state.

\subsubsection{Chaos Engineering Experiment}

To answer RQ2, we conduct chaos engineering experiments on each client.
During a chaos engineering experiment, all of the components in \chaoseth are activated. The system call error injector intercepts a specific type of system call invocation. It replaces the successful return code with an error code while the Ethereum client is connected to the rest of the blockchain distributed system.
To perform the chaos engineering experiment, recall that we use the results of the steady state analysis, see \autoref{sec:orchestrator}, and we use those of RQ1. Also, the duration of the warm-up phase is 2 hours. The pre-check, error injection, and validation phases of a chaos engineering experiment in \autoref{fig:experiments} are set to 5 minutes for each phase. The recovery phase between the error injection phase and the validation phase is set to 10 minutes in order to give a client more time for self-recovery.

\subsubsection{Resilience Benchmarking Experiment}
\label{sec:experiment-protocol-benchmarking}

RQ3 is about resilience benchmarking experiments, which aim at providing insights for end-users who wish to choose a suitable Ethereum client, based on resilience requirements. Our benchmarking experiments indicate which client has a better resilience against some system call invocation errors. 
For a resilience benchmarking experiment, we define a single, common error model for all of the target clients in order to have a sound comparison. A common error model \texttt{(s, e, $r_m$)} is one error type \texttt{(s, e, $r_c$)} that has been observed on each client in isolation. The common error rate $r_m$ is defined as the maximum value of $r_c$. The last row in  \autoref{fig:experiments} illustrates our procedure for resilience benchmarking.
Once both of the clients reach their steady state, \chaoseth conducts chaos engineering experiments using the common error model on each client. A cross-comparison is made among each client's resilience report that is generated in the chaos engineering experiment.
For example, \chaoseth injects an \texttt{ENOMEM} (insufficient memory) error into a \texttt{read} system call invocation. If client A directly crashes while client B continues to run and conducts retries after this error is injected, client B is more resilient with respect to the injected type of errors.

\subsection{Experiment Results}\label{sec:experiment-results}

\subsubsection{Steady State Analysis Experiment Results}

We perform a steady state analysis experiment on each selected Ethereum client per \autoref{sec:experiment-protocol-steadystate}. Every client is observed for two monitoring epochs of 5 hours, amounting to 10 hours in total. Within each epoch, the metrics of interest are recorded and aggregated for every 15-second monitoring interval.

In the case of GoEthereum, the client exposes $400$ metrics in total. \chaoseth analyzes all of them in order to identify proper metrics for chaos engineering experiments. As not all of the client's features are activated using the recommended configuration, $164$ out of these $400$ metrics are inactive, meaning that the value of these metrics never change, or cannot be queried. The comparison of the distributions of active metric values in two different monitoring epochs, shows evidence that $44$ metrics are statistically stable. As mentioned in \autoref{sec:steady-state-analyzer}, the steady state analyzer utilizes the Mann Whitney U test for distribution comparison. In all of the experiments, we use a p-value of $0.03$. Under a confidence level of $0.01$, the null hypothesis that the two samples are not statistically distinguishable is not rejected. In the case of Nethermind, $231$ metrics are analyzed by the steady state analyzer, $115$ metrics are inactive during the experiment. Regarding the $116$ active metrics, $55$ are statistically stable and can be used for further experiments.

\begin{figure}
    \centering
    \subfloat[GoEthereum (400 metrics in total)]{{\includegraphics[width=.45\columnwidth]{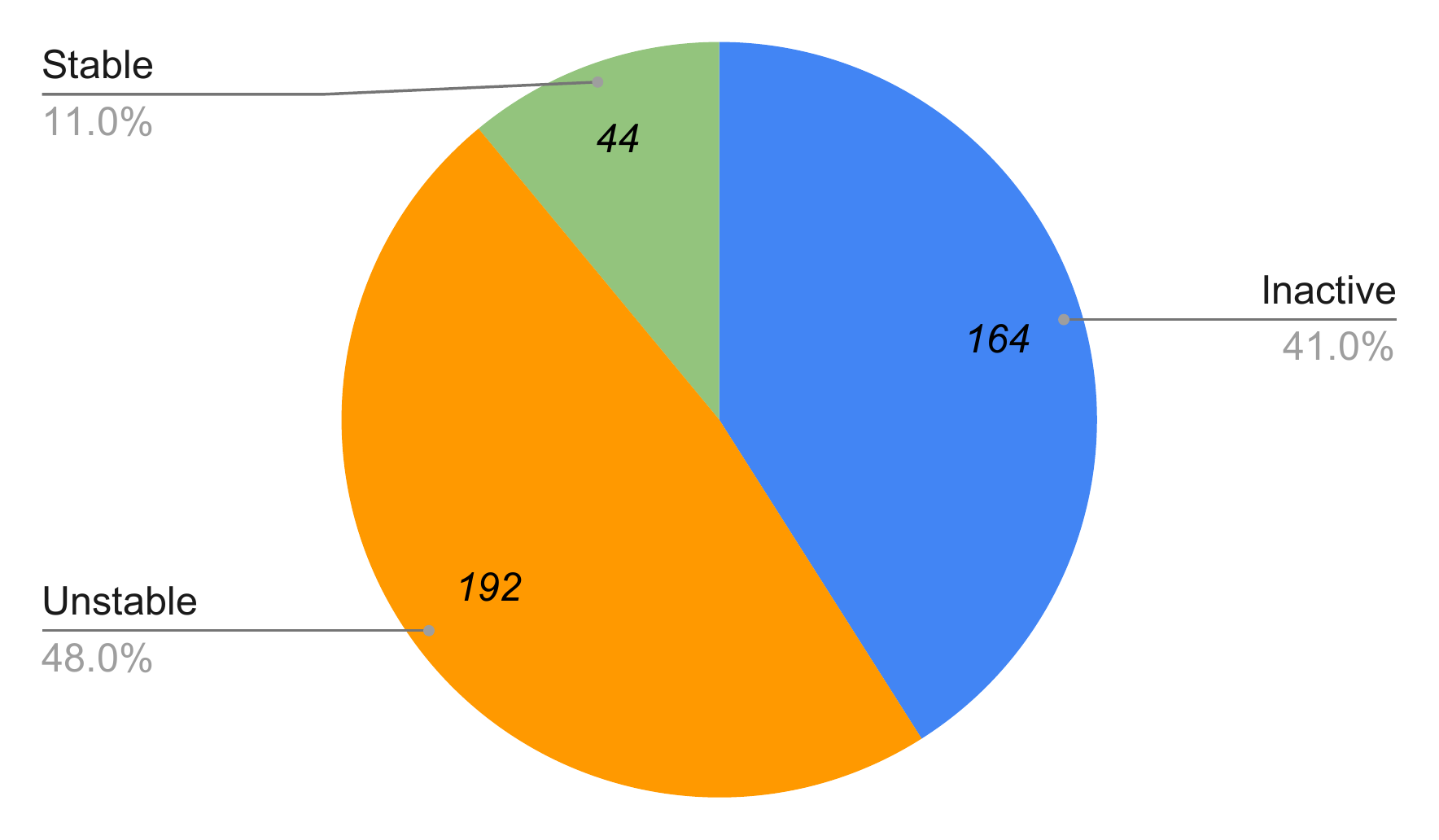}}}%
    \subfloat[Nethermind (231 metrics in total)]{{\includegraphics[width=.45\columnwidth]{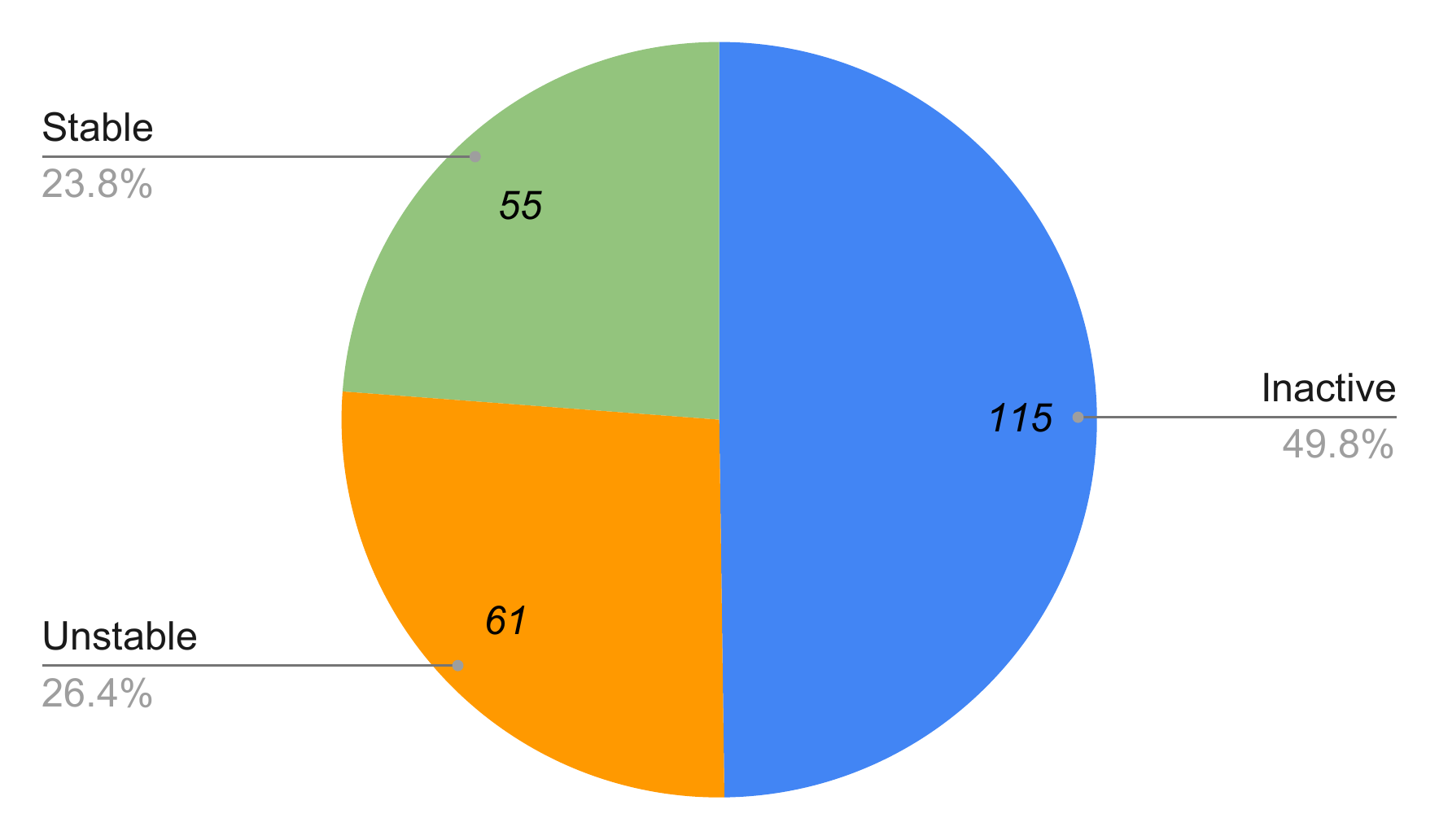}}}%
    \caption{The Distribution of Metrics in GoEthereum and Nethermind}%
    \label{fig:steady-state-analysis-distribution}%
\end{figure}

In \autoref{tab:ssa-experiment-results}, we display the evolution of metric samples during the steady state experiment. The first half of each evolution chart (in blue) is based on the data gathered during the first monitoring epoch. The second half of the chart (in red) is drawn based on the data of the second monitoring epoch. The last two columns indicate the p-values, obtained after applying the Mann–Whitney U test, to test for the similarity between the two distributions, and the result of the test. For example, the first row in \autoref{tab:ssa-experiment-results} shows that the number of account flush operations made by the GoEthereum client regularly has spikes during these two monitoring epochs.

An example where the null hypothesis is rejected at confidence level $0.01$ is metric \texttt{json.rpc.req\-uests(count/s)} in the Nethermind client. From \autoref{tab:ssa-experiment-results} the line chart in the second-to-last row also visually confirms that the metric does not evolve in the same way during the two monitoring epochs. Considering the confidence level of $0.01$, this metric is not stable enough to describe a client's steady state and thus is excluded from further experiments.

\begin{table}[tb]
\centering
\scriptsize
\caption{Samples of the Steady State Analysis Experiment Results (The colors blue and red in an evolution chart stand for the data captured in the two different monitoring epochs.)}\label{tab:ssa-experiment-results}
\begin{tabular}{lllrc}
\toprule
\textbf{Client}& \textbf{Metric Name and Unit}& \textbf{Trend}& \textbf{p-value}& \textbf{Steadiness}\\
\midrule
\multirow{6}{*}{GoEthereum}& state.snapshot.flush.account.item(count/s)& \includegraphics[width=4.5cm]{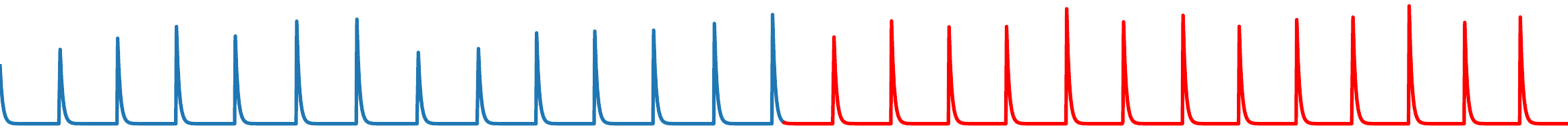}& $0.17$& √\\
& chain.inserts(operations/s)& \includegraphics[width=4.5cm]{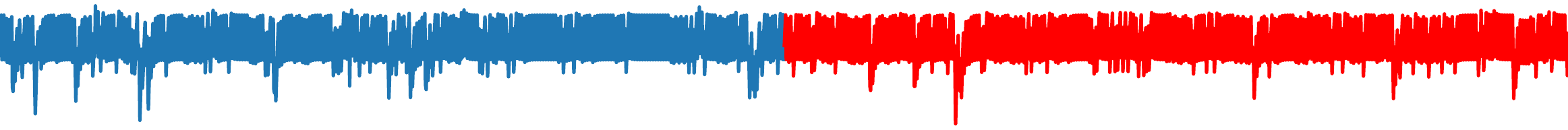}& $0.69$& √\\
& p2p.peers(count/s)& \includegraphics[width=4.5cm]{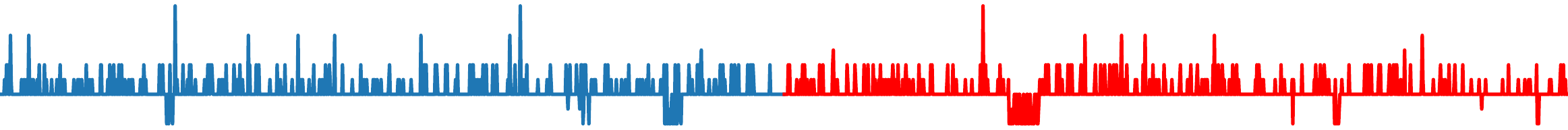}& $0.31$& √\\
& rpc.duration.all(seconds)& \includegraphics[width=4.5cm]{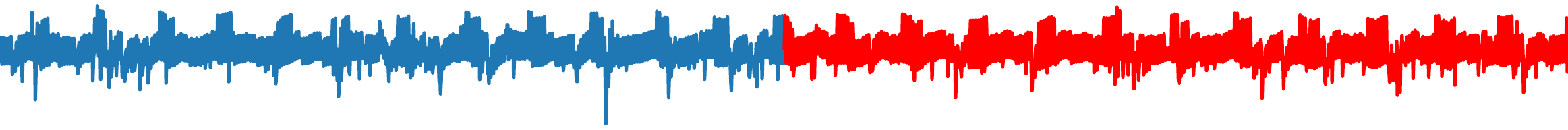}& $0.61$& √\\
& eth.db.chaindata.disk.read(bytes/s)& \includegraphics[width=4.5cm]{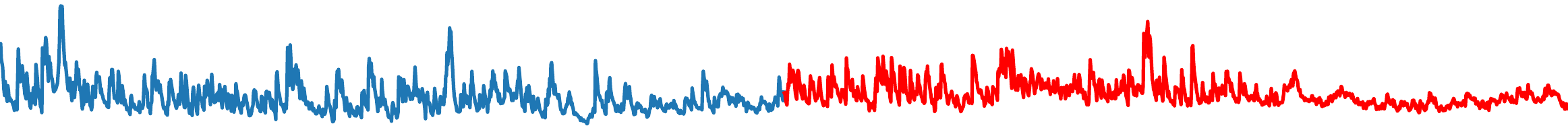}& $<0.01$& \\
& p2p.dials(count/s)& \includegraphics[width=4.5cm]{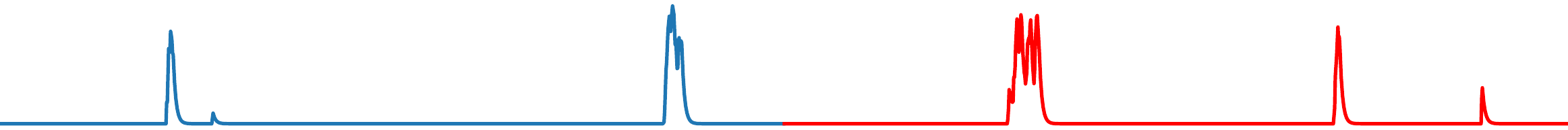}& $<0.01$& \\
\midrule
\multirow{5}{*}{Nethermind}& statuses.sent(count/s)& \includegraphics[width=4.5cm]{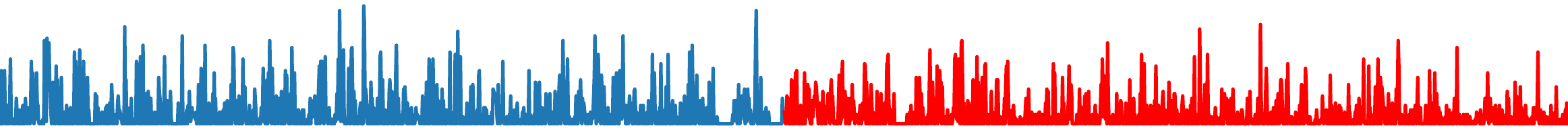}& $0.06$& √\\
& state.db.reads(bytes/s)& \includegraphics[width=4.5cm]{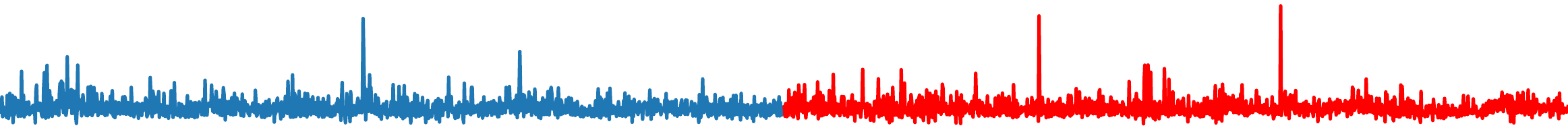}& $0.94$& √\\
& hellos.received(count/s)& \includegraphics[width=4.5cm]{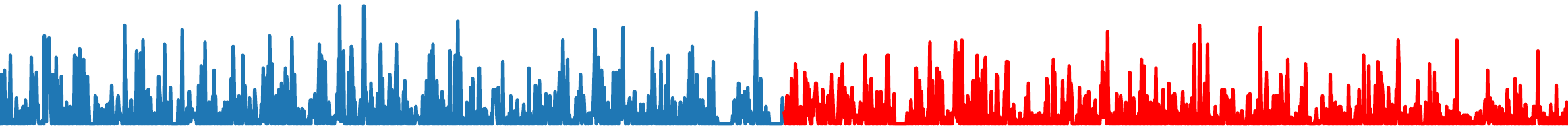}& $0.05$& √\\
& outgoing.connections(count/s)& \includegraphics[width=4.5cm]{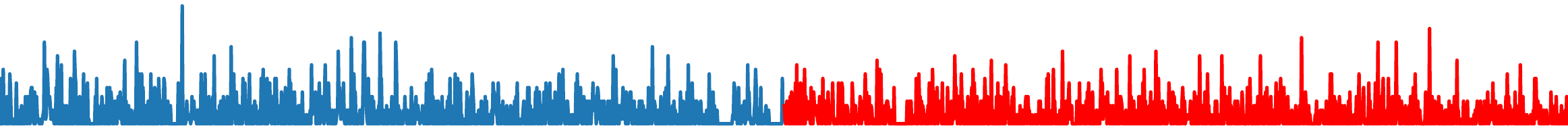}& $0.29$& √\\
& json.rpc.requests(count/s)& \includegraphics[width=4.5cm]{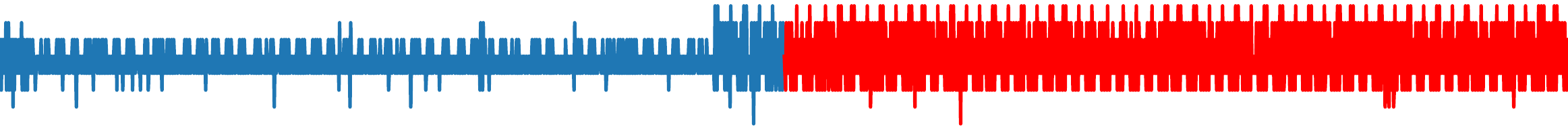}& $<0.01$& \\
& pending.transactions.added(count/s)& \includegraphics[width=4.5cm]{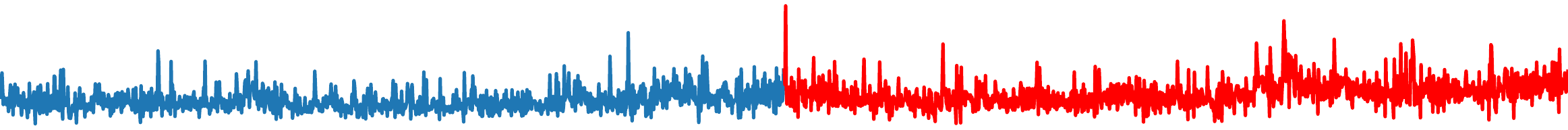}& $<0.01$& \\
\bottomrule
\end{tabular}
\end{table}

This experiment shows that not all the monitoring metrics provided by an Ethereum client are suitable to describe the client's steady state in a statically valid manner. Since the experiments are done in production, there are several factors that could affect a metric's stability. First, the node itself is not always stable. For instance, there exist other applications that take more resources from the node. Second, the network may not be stable. It is possible that the node encounters network scans or attacks every now and then \cite{Chen:SurveyOnEthSecurity}. Lastly, the behaviors of peers are different. For example, when the node randomly connects with some new peers who have different characters, a metric might be influenced.

\begin{mdframed}[style=mpdframe,nobreak=true,frametitle=Answer to RQ1]
The results of the steady state analysis experiments show that not all the monitoring metrics provided by an Ethereum client are suitable to describe the client's steady state in a statistically valid manner. \chaoseth is able to successfully identify monitoring metrics that are stable in production: 44 out of the 236 active metrics in GoEthereum, and 55 out of the 116 active metrics in Nethermind are selected for describing the steady state.
\end{mdframed}

\subsubsection{Chaos Engineering Experiment Results}

From the experiment for RQ1, we know that GoEthere-um runs with 10 different types of system calls, accumulating more than 288 million invocations in a $10$\nobreakdash-hour production run (two monitoring epochs).
Interestingly, none of the types of system call invocations has a 100\% success rate. 
We perform chaos engineering by increasing the error rates of those system calls in production.
The error rate amplification approach described in \autoref{sec:injector} produces 15 and 12 realistic error models respectively for the GoEthereum client and the Nethermind client. For each error model, we make a chaos engineering experiment, with a one-to-one mapping.

\autoref{tab:ce-experiment-results} describes the error models together with the chaos engineering experiments of the selected clients. Every row presents one error injection model, including the target system call invocation (column Syscall), the error code to be injected, and the error rate. The last five columns give the corresponding experiment result, including the total number of injected errors, the number of evaluated metrics, and the results of whether the three hypotheses ($H_N$, $H_O$, $H_R$) are verified or falsified with respect to a metric. The metrics that fail the pre-check phase are excluded from the other phases since \chaoseth considers them not stable enough for behavior comparison. When the client does not invoke a type of system call during the experiment, \chaoseth does not inject any error related to that type of system call, and the corresponding row is omitted in the table.

\begin{table}[tb]
\centering
\small
\caption{Chaos Engineering Experimental Results on the Major Ethereum Clients}
\label{tab:ce-experiment-results}
\begin{tabular}{lllrrrcrr}
\toprule
\textbf{Client}& \textbf{Syscall}& \textbf{Error Code}& \textbf{Error Rate}& \textbf{Injections}& \textbf{Metrics}& \textbf{H\textsubscript{N}}& \textbf{H\textsubscript{O}}& \textbf{H\textsubscript{R}}\\
\midrule
\multirow{12}{*}{GoEthereum}& accept4& EAGAIN& 0.6& 670& 24& √& 18& 16\\
& connect& EINPROGRESS& 0.8& 206& 4& √& 1& 1\\
& epoll\_ctl& EPERM& 0.164& 405& 24& √& 3& 0\\
& epoll\_pwait& EINTR& 0.05& 2781& 29& √& 1& 0\\
& futex& EAGAIN& 0.05& 2& -& X& -& -\\
& futex& ETIMEDOUT& 0.05& 4& -& X& -& -\\
& read& EAGAIN& 0.559& 10680& 0& √& 0& -\\
& read& ECONNRESET& 0.05& 413& 24& √& 5& 3\\
& recvfrom& EAGAIN& 0.6& 1500& 24& √& 3& 1\\
& write& EAGAIN& 0.05& 106& -& X& -& -\\
& write& ECONNRESET& 0.05& 2012& -& X& -& -\\
& write& EPIPE& 0.05& 892& -& X& -& -\\
\midrule
\multirow{10}{*}{Nethermind}& accept4& EAGAIN& 1& 1524& 51& √& 4& 2\\
& futex& EAGAIN& 0.05& 2& -& X& -& -\\
& futex& ETIMEDOUT& 0.05& 1& -& X& -& -\\
& recvfrom& EAGAIN& 0.549& 51715& 48& √& 0& -\\
& recvfrom& ECONNRESET& 0.05& 2612& 42& √& 28& 23\\
& recvmsg& EAGAIN& 1& 6588& 50& √& 2& 1\\
& sendmsg& EAGAIN& 0.05& 2277& 52& √& 16& 14\\
& sendmsg& ECONNRESET& 0.05& 1991& 47& √& 13& 10\\
& shutdown& ENOTCONN& 0.05& 82& 51& √& 2& 0\\
& unlink& ENOENT& 0.577& 40& 42& √& 2& 2\\
\bottomrule
\multicolumn{9}{p{12cm}}{
H\textsubscript{N}: `√' if the injected errors do not crash the client, otherwise `X'.\newline
H\textsubscript{O}: The number of metrics that the injected errors have a visible effect on.\newline
H\textsubscript{R}: The number of metrics that matche its steady state during the validation phase.\newline
If a hypothesis is left to be untested, it is marked as `-'.}
\end{tabular}
\end{table}

For the GoEthereum client, \chaoseth conducts 12 chaos engineering experiments. The results show that 5 out of 12 error models crash the GoEthereum client (the rows whose $H_N$ column is marked with ``X''). For the other 7 error models, 6 of them have a visible effect on the monitoring metrics (the rows whose $H_O$ column contains a non-zero value). For example, when \chaoseth uses the error model \texttt{(accept, EAGAIN, 0.6)} for experiments, $24$ metrics are stable during the pre-check phase. During the error injection phase, $18$ metrics are observed to deviate from their normal behavior. After stopping injecting the errors, $16$ out of these $18$ metrics recover to the normal state after the recovery phase. This confirms that the GoEthereum client is resilient to \texttt{EAGAIN} errors in \texttt{accept4} with respect to these 16 metrics.

Regarding the Nethermind client, there are 10 chaos engineering experiments in total (second half of \autoref{tab:ce-experiment-results}). The results show that two error models, \texttt{(futex, EAGAIN, 0.05)} and \texttt{(futex, ETIMEDOUT, 0.05)}, lead the Nethermind client to a crash. Seven error models cause a visible effect on at least one metric during the error injection phase. The error model \texttt{(recvfrom, EAGAIN, 0.549)} does not cause any impact on all of the $48$ metrics that pass the pre-check. In this case, \chaoseth does not check the $H_O$ hypothesis because no metric deviates from its steady state even during the error injection phase.

This experiment has five main outcomes, with different meanings for the Ethereum developers.

\paragraph{Crash ($H_N$=X)} The client directly crashes because of the injected errors. This is considered as a severe case: this means that an Ethereum node disappears from the distributed consensus and validation process. As the client crashes, the hypotheses $H_O$ and $H_R$ cannot be tested and are marked as `-' in \autoref{tab:ce-experiment-results}. For example, \chaoseth detects that the GoEthereum client directly crashes when an \texttt{EAGAIN} error code is injected to the system call \texttt{write}. Since error code \texttt{EAGAIN} in Linux means that the target resource is temporarily unavailable, crashing is an over-reaction, the client should consider implementing a classical retry mechanism instead of crashing directly.

\paragraph{Invisible effect ($H_N$=√ and $H_O$=0)} In some cases, there is no visible effect detected during the error injection phase. For example, the Nethermind chaos experiment using error model \texttt{(recvfrom, EAGAIN, 0.549)} reveals such a situation. In this experiment, \chaoseth injects $51715$ system call invocation errors to system call \texttt{recvfrom}. During this error injection phase, none of the $48$ metrics have an abnormal behavior. This indicates that the Nethermind client seems to be functioning normally when a system call invocation to \texttt{recvfrom} returns an \texttt{EAGAIN} error code, which can potentially signify resilience. However, we cannot exclude that the client state is corrupted in an invisible manner, because we do not have a provably perfect steady state oracle.
Since \chaoseth does not capture anything abnormal during the error injection phase, the verification of hypothesis $H_R$ is skipped.
Overall, the presence of such invisible effect cases is good with respect to consistency: if we would not perform steady state pre-checking and observability hypothesis checking, developers may falsely believe that the client state is valid according to the monitored metrics.

\paragraph{Long-term effect ($H_N$=√, $H_O$=√ and $H_R$=X)} For some of the error models, the client under experiment does not crash. However, during the error injection phase, some metrics deviate from their steady state, and do not recover after the given recovery phase. For instance, the experiment result of error model \texttt{(accept4, EAGAIN, 1)} in the Nethermind client belongs to this category. During the error injection phase, it shows that metrics \texttt{eth66get\_block\_headers\_received/s}, \texttt{local\_receive\_message\_timeout\_disconnects/s}, \texttt{process\_private\_memory/s}, and \texttt{proces\-s\_virtual\_memory/s} deviate from their normal behavior. However, after the recovery phase, only metrics \texttt{process\_private\_memory/s} and \texttt{process\_virtual\_memory/s} recover to the steady state. The other two metrics stay abnormal during the validation phase.
This means that either it takes a longer time for the client to recover from the injected errors, or that the injected errors lead the client to a stalled or corrupted state. 
Overall, such cases show that \chaoseth gives Ethereum developers insights about the timespan of recovery.

\paragraph{Resilient case ($H_N$=√, $H_O$=√ and $H_R$=√)} Certain error models do not crash the client and also cause visible evidence of resilience. After the error injection stops, the monitoring metrics recover to their steady state.
This indicates that the target client is equipped with an effective, graceful error-handling mechanism that brings the client back to normal after errors.
For example, during the chaos engineering experiment using error model \texttt{(connect, EINPROGRESS, 0.8)} in the GoEthereum client, the injected errors do not crash the client, thus the $H_N$ hypothesis holds. During the error injection phase, the metric \texttt{geth.txpool.slots.gauge/s} no longer matches the steady state. When the error injection stops, the client's behavior related to the transaction pool slots is restored during the recovery phase. During the validation phase, \chaoseth checks the metric again and confirms that \texttt{geth.txpool.slots.gauge/s} has recovered to its steady state. By looking at the client logs, we indeed confirm that the client has resumed downloading, sharing and verifying Ethereum blocks.

\begin{mdframed}[style=mpdframe,nobreak=true,frametitle=Answer to RQ2]
\chaoseth successfully conducts 12 and 10 different chaos engineering experiments respectively on the GoEthereum and the Nethermind blockchain clients. The results show that the clients have different degrees of resilience with respect to system call invocation errors. \chaoseth demonstrates that the clients crash under errors that are recoverable in theory. Since clients may crash concomitantly, this is a threat to the consensus and resilience properties of the Ethereum network from a systemic perspective. \chaoseth gives valuable insights about resilience founded on well-defined chaos engineering hypotheses.
\end{mdframed}

\subsubsection{Benchmarking Ethereum Clients}
\label{sec-result-benchmarking}
\begin{table*}[tb]
\centering
\scriptsize
\caption{Resilience Benchmarking Experiment Results}\label{tab:benchmarking-experiment-results}
\begin{tabularx}{\textwidth}{llr|rlXXX|rlXXX}
\toprule
\multicolumn{3}{c}{\textbf{Common Error Model}}& \multicolumn{5}{c}{\textbf{GoEthereum}}& \multicolumn{5}{c}{\textbf{Nethermind}}\\
\cmidrule(lr){1-3} \cmidrule(lr){4-8} \cmidrule(lr){9-13}
\textbf{Syscall}& \textbf{Error Code}& \textbf{Error Rate}& \textbf{Injections}& \textbf{Metrics}& \textbf{H\textsubscript{N}}& \textbf{H\textsubscript{O}}& \textbf{H\textsubscript{R}}& \textbf{Injections}& \textbf{Metrics}& \textbf{H\textsubscript{N}}& \textbf{H\textsubscript{O}}& \textbf{H\textsubscript{R}}\\
\midrule
accept4& EAGAIN& 1& 736& 35& √& 9& 2& 1344& 49& √& 6& 3\\
futex& EAGAIN& 0.05& 4& -& X& -& -& 2& -& X& -& -\\
futex& ETIMEDOUT& 0.05& 3& -& X& -& -& 1& -& X& -& -\\
recvfrom& EAGAIN& 0.6& 1257& 0& √& 0& -& 26691& 48& √& 0& -\\
\bottomrule
\end{tabularx}
\end{table*}

We cannot strictly compare the considered clients based on the results of RQ2, because the error models are different. To overcome this, we have introduced in \autoref{sec:experiment-protocol-benchmarking} the idea of testing the clients under a meaningful common error model. \chaoseth identifies $4$ common error models for the selected clients. The results of this resilience benchmarking experiment are summarized in \autoref{tab:benchmarking-experiment-results}. Each row in the table presents the verification of the three hypotheses for both clients, according to a set of client metrics. Only the metrics that pass the pre-check phase are selected for hypothesis verification. This table is interesting in the following three aspects.

First of all, regarding the $H_N$ hypothesis (absence of crash), the results show that both the GoEthereum client and the Nethermind client crash under the same specific error models. These two clients are crashed by \texttt{futex} system call invocation errors with codes \texttt{EAGAIN} and \texttt{ETIMEDOUT}. Overall, there is no client which is absolutely more robust than the other with respect to crashing.

Second, focusing on the $H_O$ hypothesis (observability), when the error model \texttt{(accept4, EAGAIN, 1)} is used for experiments, both the GoEthereum client and the Nethermind client are observed to have abnormal behavior with respect to metrics. For the GoEthereum client, $9$ metrics become abnormal during the error injection phase. Regarding the Nethermind client, $6$ metrics deviate from the steady state. This is evidence that the metrics capture the client's internal state, and that not all clients have the same observability.

Third, considering the $H_R$ hypothesis, \chaoseth successfully identifies resilient cases for the two Ethereum clients.
\chaoseth shows that the GoEthereum client is resilient to error model \texttt{(accpet4, EAGAIN, 1)} with respect to metrics \texttt{geth.p2p.peers.gauge/s} and \texttt{geth.txpool.reheap.timer/s}. For the same error model, the Nethermind client is resilient with respect to metrics \texttt{nethermind\_mo\-d\_exp\_precompile/s}, \texttt{nethermind\_state\_db\_reads/s}, and \texttt{nethermind\_useless\_peer\_disc\-onnects/s}.
As opposed to toy examples with perfect oracles, assessing behavior of real-world software through monitoring yields multiple shades of resilience.

\begin{mdframed}[style=mpdframe,nobreak=true,frametitle=Answer to RQ3]
\chaoseth identifies $4$ common error models for resilience benchmarking. The results show that neither of the clients is consistently more resilient than the other. For \texttt{futex} system call invocation errors, both the GoEthereum client and the Nethermind client crash. When injecting \texttt{EAGAIN} errors in system call invocations to \texttt{accept4}, the GoEthereum client has $9$ metrics that deviate from the steady state but only $2$ of them recover. The Nethermind client has $6$ metrics that are affected by the injected errors, $3$ out of which are able to recover. In both cases, given the nature of the injected errors (\texttt{EAGAIN} and \texttt{ETIMEDOUT}), the clients could implement a retry mechanism for better resilience.
\end{mdframed}

\section{Discussion}\label{sec:discussion}

\subsection{Threats to Validity}
Per our methodology, during a steady state analysis experiment, a set of statistically stable metrics is identified. A corresponding threat to the internal validity relates to the selection criterion based on a Mann–Whitney U test. The threshold on the p-value may impact the selection results and subsequent outcome of the chaos engineering experiments. Another thread to the internal validity is that all the experiments are done in the environment mentioned in \autoref{sec:experiment-protocol}. 
The generalizability of our results can be strengthened in two ways: First, in this work, we evaluate the Ethereum clients under a synchronization workload. However, different workloads, such as performing RPC invocations, would most likely vary the behavioral metrics and consequently the captured steady state.
Second, since system calls are implemented differently by operating systems, changing the operating system for chaos experiments may have an impact on the results.

A threat to the construct validity is that, in the chaos engineering experiments, errors are injected by replacing a successful return code of a system call with an error return code according to a pre-defined error model. However, the system call itself is still executed according to the eBPF execution model, potentially modifying the program's state as originally intended. It is possible that abnormal behavior observed under error injection is due to the inconsistency between the returned error code and the internal state of the program, and not to the returned error code alone.

\subsection{Ethical Considerations}

The more diverse the exchanged messages are, the better it is for \chaoseth to detect interesting behavior during fault injection experiments. From this perspective, it is beneficial to apply \chaoseth directly to the main network of Ethereum, per the core principles of chaos engineering. At the same time, the possibility of harming the network has to be evaluated. First of all, we make sure that \chaoseth only perturbs a single Ethereum client. Secondly, during a fault-injection experiment the perturbed client only downloads and verifies blocks, making it less likely to send out malformed messages to other peers. Finally, we have made the Ethereum Foundation aware of this research project, and received their endorsement for this line of the software reliability research.

\subsection{Applicability to Other Clients}

As \chaoseth does chaos engineering experiments at the system call invocation level, theoretically any Ethereum client implementation that runs on top of a Linux operating system can be evaluated by \chaoseth. As long as the operating system has the eBPF module installed, \chaoseth is able to monitor the system call invocations and inject specific errors. No specific changes are required in the client for attaching \chaoseth. At the conceptual level, our methodology is not limited to Ethereum or EVM compatible blockchains, it can be applied to other blockchains as well.

\section{Conclusion}\label{sec:conclusion}
In this paper, we have presented a novel chaos engineering framework called \chaoseth. It actively injects system call invocation errors into Ethereum client implementations to assess their resilience in production. Our experiments show that \chaoseth is effective to detect system call related error-handling weaknesses and strengths, ranging from direct crashes to full resilience in two of the most popular Ethereum clients, GoEthereum and Nethermind. As a direction for future research, it is promising to investigate the multiple kinds of resilience improvement in a blockchain client: at the operating system level, at the level of standard libraries, and in the client's code, in order to achieve ultra-reliable blockchain systems.

\begin{acks}
This work was supported by the Wallenberg Artificial Intelligence, Autonomous Systems and Software Program (WASP) funded by Knut and Alice Wallenberg Foundation, and by the Swedish Foundation for Strategic Research (SSF). Some experiments were performed on resources provided by the Swedish National Infrastructure for Computing (SNIC).
\end{acks}
\bibliographystyle{ACM-Reference-Format}
\bibliography{references}

\end{document}